\documentclass[useAMS,usenatbib]{mn2e}

\usepackage[applemac]{inputenc}
\usepackage{epsfig}
\usepackage{dcolumn}
\usepackage[fleqn]{amsmath}
\usepackage{bm}

\title[Baryons and Weibel instability in GRBs]{Baryon loading and the Weibel instability in gamma-ray bursts}
\author[M. Fiore et al.]
  {M. Fiore,$^1$\thanks{E-mail: m.fiore@ist.utl.pt} L. O. Silva,$^1$\thanks{E-mail: luis.silva@ist.utl.pt}
  C. Ren,$^2$ M. A. Tzoufras$^3$ and W. B. Mori$^{3,4}$\\
  $^1$GoLP/Centro de Física dos Plasmas, Instituto Superior Técnico, 1049-001 Lisboa, Portugal\\
  $^2$Department of Mechanical Engineering, Department of Physics and Astronomy,\\ and Laboratory for Laser Energetics, University of Rochester, Rochester, NY 14627, USA\\
  $^3$Department of Electrical Engineering, University of California, Los Angeles, CA 90095, USA\\
  $^4$Department of Physics and Astronomy, University of California, Los Angeles, CA 90095, USA}
\date{\today}

\pagerange{\pageref{firstpage}--\pageref{lastpage}} \pubyear{2006}

\def\LaTeX{L\kern-.36em\raise.3ex\hbox{a}\kern-.15em
    T\kern-.1667em\lower.7ex\hbox{E}\kern-.125emX}

\begin{document}

\label{firstpage}

\maketitle

\begin{abstract}
The dynamics of two counter-streaming electron-positron-ion unmagnetized plasma shells with zero net charge is analyzed in the context of magnetic field generation in GRB internal shocks due to the Weibel instability. The effects of large thermal motion of plasma particles, arbitrary mixture of plasma species and space charge effects are taken into account. We show that, although thermal effects slow down the instability, baryon loading leads to a non-negligible growth rate even for large temperatures and different shell velocities, thus guaranteeing the robustness and the occurrence of the Weibel instability for a wide range of scenarios.
\end{abstract}


\section{Introduction}

One of the unsolved problems in the astrophysics of extreme events is the magnetic field generation in gamma-ray bursts (GRBs). It is almost universally accepted the hypothesis that the internal and the external shocks in relativistic ejecta represent the most likely mechanism for $\gamma$-ray production, with synchrotron emission as the main radiation process in GRBs \citep{sar97}, thus indicating the presence of strong particle acceleration and near-equipartition magnetic fields. Although the latter has been confirmed by synchrotron radiation measurements, the origin of the required magnetic field is still under strong debate \citep{cob03,wax03}. The Weibel instability \citep{wei59}, which has received an increasing interest in recent years, seems to be a good candidate to explain magnetic fields in GRBs \citep{med99,med05}, as confirmed by three dimensional particle-in-cell simulations \citep{sil03,fre04,nis03}. 

In this paper, we examine the dynamics of two counter-streaming electron-positron-ion unmagnetized plasma shells with zero net charge, scenario that applies to the internal shocks of GRBs. Using relativistic kinetic theory we focus our discussion on the growth rate ($\Gamma$) of the electromagnetic beam-plasma instability or Weibel instability, since it provides an estimate of the maximum magnetic field generated by the instability. Indeed, the saturation level $b_{sat}$ of the magnetic field is related to $\Gamma$ according to $b_{sat}\propto\Gamma^2$ \citep{med99,dav72,yan94}. A similar procedure was used to examine the magnetic field generation due the Weibel instability in GRB external shocks \citep{wie04}. In previous analytical works on the Weibel instability in GRBs, the relativistic plasma is assumed cold and only purely electromagnetic transverse modes are excited; moreover, no arbitrary mixture of plasma components has been considered. In this work, we employ relativistic kinetic theory to include thermal effects, with arbitrary temperatures, space charge effects and an arbitrary mixture of leptons and baryons.

It is well known that thermal effects can strongly suppress and even shutdown the current filamentation instability \citep{sil02}. Furthermore, a non relativistic analysis \citep{tzo06} has shown that the component of the electric field {\boldmath$E$} along the wave vector {\boldmath$k$} can also be important for two inter-penetrating current beams having different temperatures. The different rate at which these current beams pinch leads to density filaments, thus creating a space charge which reduces the growth rate and causes ions to respond. It is important to consider such effects because the relativistic shells ejected by the central engine have not only a spread in the bulk velocities but should also show the presence of an arbitrary mixture of electrons, positrons and protons, with varying temperature from shell to shell \citep{ree92,laz99,pan99}.

We have studied several scenarios by changing the properties of the colliding shells (e.g. particle number density, temperature and bulk velocity of the shell) in order to have a more complete analysis of the phenomena that arise in these scenarios. We show that, although thermal effects slow down the instability, the thermal spread does not shutdown the Weibel instability. The presence of ions becomes crucial to sustain the instability when shells with different velocities are considered. It is shown that baryon loading leads to a non-negligible growth rate by lowering the threshold predicted for purely electromagnetic waves in an electron-positron plasma \citep{sil02,fon03}, thus guaranteeing the robustness and the occurrence of the instability even for colliding shells with large temperatures and different mixtures of plasma components. The threshold for the occurrence of the Weibel instability, including relativistic and space charge effects, is also discussed.

This paper is organized as follows. The theoretical model employed in our analysis is described in Section 2. In Section 3 we present a discussion of the numerical results regarding the growth rate of the Weibel instability and the threshold for its occurrence. Finally, Section 4 summarizes our main conclusions.

\section{Theoretical model}

To motivate our discussion, let us consider a simple model for two counter-streaming plasma shells ejected by the central engine of the burst \citep{laz99}, with respect to the centre of mass (CM) reference frame of the two-shell system \citep{med99}. Each shell is characterized by an arbitrary mixture of electrons, positrons and protons, a bulk Lorentz factor $\gamma$ and a bulk thermal Lorentz factor $\gamma_{th0}=(1-\beta_{th0}^2)^{-1/2}$, being $\beta_{th0}$ the thermal velocity. The total charge in each shell and the momentum in the CM reference frame are always conserved.

In order to obtain the dispersion relation for waves propagating in the $z$ direction with wave vector {\boldmath$k$}$=k${\boldmath$e_z$}, we use the relativistic Vlasov equation and Maxwell's equations, following the technique outlined by \citet{sil02}, yielding:
\begin{equation}
\begin{split}
\label{eq:1}
&\left(\omega^2-k^2c^2+C_{xxz}\right)\left[\left(\omega^2-k^2c^2+C_{yyz}\right)\left(\omega^2+D_z\right)-D_yC_{zyz}\right]\\
&-C_{xyz}\left[C_{yxz}\left(\omega^2+D_z\right)-D_yC_{zxz}\right]\\
&+D_x\left[C_{yxz}C_{zyz}-\left(\omega^2-k^2c^2+C_{yyz}\right)C_{zxz}\right]=0
\end{split}
\end{equation}
with
\begin{equation}
\begin{split}
\label{eq:2}
&C_{lmn}=\sum_j\omega^2_{p0j}m_j\int \mathrm{d} \bmath{p} \frac{\omega v_l}{\omega-kv_z}\\
&\times\left\{ \left(1-\frac{kv_z}{\omega} \right)\partial_{p_m}+\frac{kv_m}{\omega}\partial_{p_n}\right\}F_{0j},
\end{split}
\end{equation}

\begin{equation}
\label{eq:3}
D_l=\sum_j\omega^2_{p0j}m_j\int \mathrm{d} \bmath{p} \left\{\frac{\omega v_l}{\omega-kv_z}\partial_{p_z}\right\}F_{0j},
\end{equation}
where the $j$ component of the plasma is described by the corresponding unperturbed normalized distribution function $F_{0j}$, the unperturbed density $n_{0j}$, the mass $m_j$ and the plasma frequency $\omega_{p0j}$. For purely electromagnetic waves propagating with the same wave vector {\boldmath$k$}$=k${\boldmath$e_z$} \citep{sil02}, only $C_{xxz}$, $C_{yyz}$, $C_{xyz}$, $C_{yxz}$ appear in the linear dispersion relation; the other terms in (\ref{eq:1}) are due to the inclusion of space charge effects, i.e. {\boldmath$k$}${\cdot}${\boldmath$E$}$\not=0$. Equations (\ref{eq:1}), (\ref{eq:2}) and (\ref{eq:3}) provide a starting point for the analysis of the Weibel instability in the relativistic regime including kinetic effects. Calculation of (\ref{eq:2}) and (\ref{eq:3}) for a general distribution function is, in general, quite difficult and only possible in closed analytical form for a reduced number of distribution functions. We choose a waterbag distribution, given by $f_{0j}=n_{0j}F_{0j}$, with $F_{0j}=\left[1/\left(2p_{z0j}\right)\right]\delta\left(p_{x}-p_{x0j}\right)\delta\left(p_{y}\right)\left\{\Theta\left(p_{z}+p_{z0j}\right)-\Theta\left(p_{z}-p_{z0j}\right)\right\}$, where $p_{x0j}$ is the momentum in the $x$ direction and $p_{z0j}$ is the momentum thermal spread in the $z$ direction of the $j$ plasma component, while $\Theta(x)$ is the Heaviside step function. With this choice of the distribution function, the dispersion relation describes the Weibel instability in the relativistic warm fluid limit, for conditions far from the kinetic regime \citep{tzo06}. Such distribution function describes a cold plasma shell in the $x$ and $y$ directions, propagating along the $x$ direction, with a thermal spread along the $z$ direction. Due to the form of the distribution function, the $p_m$-, $ p_n$- and $p_z$-integrations required in (\ref{eq:1}) can be carried out in closed analytical form. We use the subscripts "1" ("2") for the shell moving in the +{\boldmath{$e_x$}}-direction (-{\boldmath{$e_x$}}-direction) and we assume, without loss of generality, that the shell-1 is faster than the shell-2; charge neutrality for each shell is also assumed.

After normalizing time to $1/\omega_{p0e^-,1}$ (where $\omega_{p0e^-,1}^2=4\pi n_{e^-0,1}e^2/m_{e^-}$) and space to the collisionless skin depth $c/\omega_{p0e^-,1}$, and performing the integration over the distribution function (note that $C_{xyz}=C_{yxz}=C_{zyz}=D_y=0$ and $C_{zxz}=D_x$), the dispersion relation (\ref{eq:1}) can be simplified to 
\begin{equation}
\label{eq:4}
\left(\omega^2-k^2+\left[C_{xxz}\right]\right)\left(\omega^2+\left[D_z\right]\right)-\left[D_x\right]^2=0
\end{equation}
where the terms in square brackets correspond to the coefficients from (\ref{eq:2}) and (\ref{eq:3}), now given by:
\begin{subequations}
\label{eq:7}
\begin{equation}
\left\lbrack C_{xxz}\right\rbrack=-\left\{\mathcal{F}+\sum_{s,j=1,1}^{2,3}\left\lbrack\frac{n_{0j,s}}{m_{j}}\right\rbrack\frac{1}{\gamma_{0j,s}}\frac{k^2\beta_{0j,s}^2}{\omega^2-k^2\beta_{th0j,s}^2}\right\}
\end{equation}
\begin{equation}
\left\lbrack D_z\right\rbrack=-\left\{\sum_{s,j=1,1}^{2,3} \left\lbrack\frac{n_{0j,s}}{m_{j}}\right\rbrack\frac{1}{\gamma_{0j,s}}\frac{1}{\omega^2-k^2\beta_{th0j,s}^2}\right\}\omega^2
\end{equation}
\begin{equation}
\left\lbrack D_x\right\rbrack=-\left\{\sum_{s,j=1,1}^{2,3}\left\lbrack\frac{n_{0j,s}}{m_{j}}\right\rbrack\frac{1}{\gamma_{0j,s}}\frac{k\beta_{0j,s}}{\omega^2-k^2\beta_{th0j,s}^2}\right\}\omega,
\end{equation}
\end{subequations}
where the sum in $s$ is over the two shells and the sum in $j$ is over the three plasma components, and where $\beta_{0j}$ is the velocity along the $x$ direction and $\beta_{th0j}$ is the perpendicular thermal velocity of the $j$ plasma component, with the standard definitions $\gamma_{0j}=\left(1-\beta_{0j}^2-\beta_{th0j}^2\right)^{-1/2}$, $u_{0j}=\gamma_{0j}\beta_{0j}$, complemented by the definitions $\left(\overline{1/\gamma_{j}}\right)=\int \mathrm{d}\bmath{p}\left(F_{0j}/\gamma\right)$, and with
\begin{equation}
\label{eq:22}
\mathcal{F}=\sum_{s,j=1,1}^{2,3} \left\lbrack\frac{n_{0j,s}}{m_{j}}\right\rbrack\left[\left(\frac{\overline{1}}{\gamma_{j,s}}\right)-\frac{1}{\gamma_{0j,s}}\frac{u_{0j,s}^2}{1+u_{0j,s}^2}\right].
\end{equation}
In order to obtain the growth rate of the instability, one has to find the roots of the dispersion relation (\ref{eq:4}) and consider the largest positive imaginary part. Due to its complexity, a fully analytical solution leads to cumbersome expressions; therefore (\ref{eq:4}) will be solved numerically. 

However, it is possible to study the threshold condition for the Weibel instability, since near the threshold $|\omega^2|\ll1$. The dispersion relation presents unstable solutions whenever $\omega^2<0$, which, from (\ref{eq:4}), (\ref{eq:7}) and (\ref{eq:22}) yields
\begin{equation}
\label{eq:8}
\mathcal{G-F}>k^2+\frac{\mathcal{I}^2}{k^2+\mathcal{H}},
\end{equation}
where
\begin{subequations}
\label{eq:19}
\begin{equation}
\mathcal{G}=\sum_{s,j=1,1}^{2,3}\left\lbrack\frac{n_{0j,s}}{m_{j}}\right\rbrack\frac{1}{\gamma_{0j,s}}\frac{\beta_{0j,s}^2}{\beta_{th0j,s}^2}
\end{equation}
\begin{equation}
\mathcal{H}=\sum_{s,j=1,1}^{2,3}\left\lbrack\frac{n_{0j,s}}{m_{j}}\right\rbrack\frac{1}{\gamma_{0j,s}}\frac{1}{\beta_{th0j,s}^2}
\end{equation}
\begin{equation}
\mathcal{I}=\sum_{s,j=1,1}^{2,3}\left\lbrack\frac{n_{0j,s}}{m_{j}}\right\rbrack\frac{1}{\gamma_{0j,s}}\frac{\beta_{0j,s}}{\beta_{th0j,s}^2},
\end{equation}
\end{subequations}
thus generalizing the threshold condition derived by \citet{sil02}.
The maximum unstable wavenumber, $k_{max}$, is then given, from (\ref{eq:8}) by
\begin{equation}
\begin{split}
\label{eq:11}
&k_{max}^2=\frac{-\left(\mathcal{H-G+F}\right)}{2}\\
&+\frac{\sqrt{\left(\mathcal{H-G+F}\right)^2-4\left(\mathcal{I}^2-\mathcal{H}\left(\mathcal{G-F}\right)\right)}}{2}
\end{split}
\end{equation}
which corresponds to a real solution whenever
\begin{equation}
\label{eq:9}
\mathcal{G-F}>\frac{\mathcal{I}^2}{\mathcal{H}}
\end{equation}
thus establishing the condition for the existence of unstable Weibel modes.
It is important to point out that the last terms on the right-hand side of (\ref{eq:8}) and (\ref{eq:9}) represent the contribution of the space charge effects. 

Space charge effects arise from the filament pinching due to the Weibel instability, which establishes a charge imbalance in the region surrounding each filament, thus leading to an electric field parallel to the wave vector {\boldmath$k$} (i.e. {\boldmath$k$}${\cdot}${\boldmath$E$}$\not=0$). This (first order) longitudinal electric field component tries to prevent pinching, making more difficult the occurrence of the Weibel instability. Without the space charge term, the threshold is equivalent to that obtained by assuming a purely transverse electromagnetic mode \citep{sil02}. Since the space charge term is non-negative, it slows down the growth of the Weibel instability, and it raises the instability threshold. Space charge effects are cancelled out by shells having the same density, the same velocity and the same temperature. 

To examine the different contribution to the space charge effects of the species we are taking into account in our model, some straightforward calculations can be done. In the limit of an almost cold plasma ($u_{th0j}\ll1$), relativistic shells ($u_{0j}\gg1$) and using $\beta_{th0e^+}\approx\sqrt{M}\beta_{th0i}$, it is easy to verify by comparison that, for a scenario with two colliding shells having the same density,  the space charge terms for ions and positrons are of the same order, i.e. $\left(\mathcal{I}_i^2/\mathcal{H}_i\right)/\left(\mathcal{I}_{e^+}^2/\mathcal{H}_{e^+}\right)\approx1$: baryons and positrons affect in the same way the threshold of the Weibel instability in what solely concern space charge effects. For scenarios of our interest, the space charge effect term for a cold plasma is cancelled for shells moving with the same velocity \citep{dav04}.    

If we neglect space charge effects and keep the previous approximations (almost cold and relativistic plasma), from (\ref{eq:9}) we recover the condition in \citet{sil02}, $\mathcal{G-F}>0$, which with the definition $\Delta_j\approx\left\lbrack\frac{n_{0j}}{m_{j}}\right\rbrack\frac{1}{\gamma_{0j}}\left(\frac{1}{\beta_{th0j}^2}-1\right)$, is written as
\begin{equation}
\label{eq:16}
\sum_{s,j=1,1}^{2,3}\Delta_{j,s}>0.
\end{equation}
The contribution $\Delta_j$ of each species is easily identified. Comparison of the contribution to (\ref{eq:16}) of baryons and positrons at the same temperature yields
\begin{equation}
\label{eq:18}
\frac{\Delta_i}{\Delta_{e^+}}\approx\frac{1-\beta_{th0i}^2}{1-\beta_{th0e^+}^2}.
\end{equation}
Since $\beta_{th0i}<\beta_{th0e^+}$ ($\beta_{th0e^+}\approx\sqrt{M}\beta_{th0i}$) the ratio (\ref{eq:18}) is greater than 1: baryons are more "efficient" than positrons to guarantee the occurrence and the robustness of the Weibel instability when the temperature increases. What is important for the occurrence of the instability is not the species temperature, but the species thermal velocity $\beta_{th0j}$ instead. We will explore this aspect in the next section.

\section{Discussion}

To assure charge neutrality in each shell, we consider an electron density $n_{0e^-,s}$, a positron density $n_{0e^+,s}$ with $\alpha_s=n_{0e^+,s}/n_{0e^-,s}$, and the consequent ion density $n_{0i,s}=\left(1-\alpha_s\right)n_{0e^-,s}$; the subscript $s$ denotes the reference shell. Our calculations are performed in the CM reference frame, which naturally imposes the condition that the total linear momentum of the system of two shells in this frame is zero. The momentum conservation along the $x$ direction yields $\left|v_{0,1}\right|\left(\sum_{j=1}^3\gamma_{0j}m_{0j}\right)_1=\left|v_{0,2}\right|\left(\sum_{k=1}^3\gamma_{0j}m_{0k}\right)_2$, where $j,k=1,2,3$ stand for electrons, positrons and ions respectively. Since all particle species in each shell have the same mean velocity (i.e. the velocity of their own shell), the latter is independent of the species considered. The momentum conservation establishes limiting values for the features of the plasma shells in the CM reference frame, and for the parameters describing the shells. We define the ratio of the absolute values of the shell velocities as $V=\frac{\left(\sum_{j=1}^3\gamma_{0j}m_{0j}\right)_1}{\left(\sum_{k=1}^3\gamma_{0k}m_{0k}\right)_2}=\frac{|v_{0,2}|}{|v_{0,1}|}$. Since we assume that the shell-1 is faster than the shell-2, the value of $V$ can vary within the range $0\leq V\leq1$. Denoting $M=m_i/m_e$ the ion and electron mass ratio and $\Lambda_s^\pm=\gamma_{0e^-,s}\pm\gamma_{0i,s}M$, it is then straightforward to obtain an expression for the ratio of the positron and electron densities in the shell-2, which reads $\alpha_2=\frac{1}{V}\frac{\Lambda_1^++\alpha_1\Lambda_1^-}{\Lambda_2^-}-\frac{\Lambda_2^+}{\Lambda_2^-}$.
Since $0\leq\alpha_2\leq1$, the value of $\alpha_1$ can only vary in the range:
\begin{equation}
\label{eq:15}
\frac{\Lambda_2^+}{\Lambda_1^-}V-\frac{\Lambda_1^+}{\Lambda_1^-}\leq\alpha_1\leq\frac{\Lambda_2^++\Lambda_2^-}{\Lambda_1^-}V-\frac{\Lambda_1^+}{\Lambda_1^-}.
\end{equation}
We observe that, according to limitations in (\ref{eq:15}), $\alpha_1$ can assume all values between 0 and 1 only when the shells have the same velocity (i.e. $V=1$) and the same temperature (i.e. $\Lambda_1^\pm=\Lambda_2^\pm$ when $V=1$), or equivalently when $\alpha_1=\alpha_2$.

As typical parameters for the colliding shells in GRB scenarios, we follow \citet{med99} and \citet{pir99}, who suggest a bulk Lorentz factors $\gamma_{int}\approx4$, and initial electron thermal Lorentz factors $\overline{\gamma}_{th0e,in}\approx2$. Since $\gamma_{th0j}=\left(1-\beta_{th0j}^2\right)^{-1/2}$ and $\gamma_{0j}=\left(1-\beta_{0j}^2-\beta_{th0j}^2\right)^{-1/2}$, the perpendicular temperature of the shells is then obtained from the energy-momentum tensor as $T_\perp=\int d{\textbf{\emph{p}}}\,p_{zj}v_{zj}F_{0j}$, which for our choice of $F_{0j}$ is $\frac{k_BT_{\perp,s}}{m_jc^2}=\frac{\gamma_{0j,s}}{2}\left[1+\frac{1-\beta^2_{th0j,s}}{2\beta_{th0j,s}}\ln\left(\frac{1-\beta_{th0j,s}}{1+\beta_{th0j,s}}\right)\right]$ \citep{sil02}. We always consider that all particles contained in the same shell have the same temperature, whereas different shells can have distinct temperatures; the shell electron density ratio is indicated by $\eta=n_{e^-0,1}/n_{e^-0,2}$. Three different mixtures of plasma components are examined: i) electron-positron plasma, which means that $\alpha_1=1$, ii) electron-positron-ion plasma, with $\alpha_1=0.5$, iii) electron-ion plasma, such that $\alpha_1=0$. 
\begin{figure}
\epsfig{file=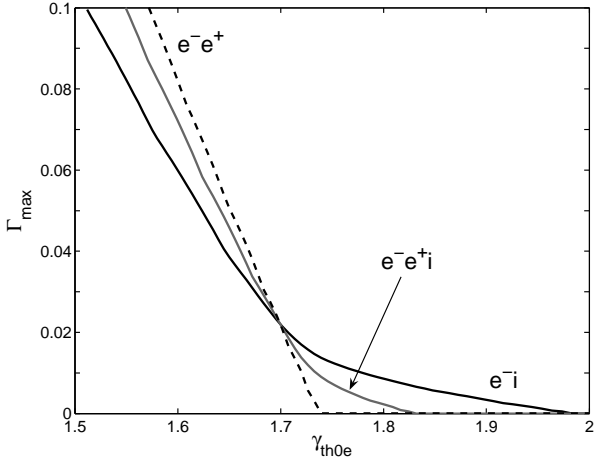,width=8cm}
\caption{Dependence of the maximum growth rate of the Weibel instability on the electron thermal Lorentz factor for shells having same velocity ($p_{x0}=2$), same temperature and same electron density, for different mixtures of plasma components: $\alpha_{1,2}=1$ (dashed line),  $\alpha_{1,2}=0.5$ (solid gray line),  $\alpha_{1,2}=0$ (solid black line).  
\label{Fig. 1}}
\end{figure}

When the colliding shells have the same velocity (i.e. $V=1$, with $p_{x0}=2$ such that $\gamma_{int}\approx4$), the same electron density (i.e. $\eta=1$) and the same temperature, the space charge effects are cancelled due to the form of the space charge term [see (\ref{eq:9})]. The dependence of the maximum growth rate of the Weibel instability on the electron thermal Lorentz factor is shown in Fig. \ref{Fig. 1}; for this scenario, $0\leq\alpha_1=\alpha_2\leq1$. As expected, the maximum growth rate decreases as temperature increases, until the Weibel instability shutdowns for all plasma component mixtures. The ion presence is fundamental to maintain the occurrence of the instability for larger temperatures than the shutdown temperature for electron-positron plasma (corresponding to $\gamma_{th0e}\approx1.75$). As the baryon component increases in the plasma mixture, the shutdown temperature for the Weibel instability to occur also increases. This result confirms the estimate of the previous section about role of baryons to sustain unstable scenarios for higher temperatures [cf. (\ref{eq:18})].

\begin{figure}
\epsfig{file=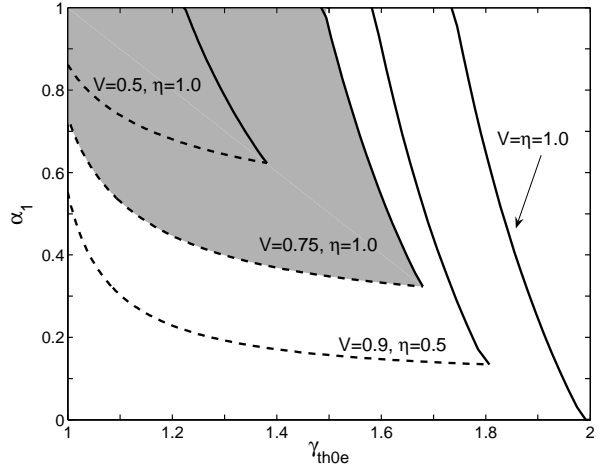,width=8cm}
\caption{Unstable area ($p_{x0}=2$) for different scenarios. Solid lines represent the threshold from (\ref{eq:9}); dashed lines represent the limits from momentum conservation [see (\ref{eq:15})]. The gray area indicates the unstable area for $V=0.75$ and $\eta=1$. 
\label{Fig. 5}}
\end{figure}
A mean to analyze a large number of the possible scenarios for the occurrence of the Weibel instability for two counter-steaming shells having same temperature, depending on the parameters $V$ and $\eta$, is provided by Fig. \ref{Fig. 5}. Here, the threshold for the occurrence of the Weibel instability described by (\ref{eq:9}) is represented by solid lines; the points below of these lines correspond to unstable scenarios. The dashed line represents the lower limit established by (\ref{eq:15}) (constraint associated with CM reference frame). The negative slope of these curves is a clear signature of the role played by baryons ($\alpha_1<1$), which allow the occurrence of the instability for larger temperatures than those for mixtures of leptons only ($\alpha_1=1$).     

\begin{figure}
\epsfig{file=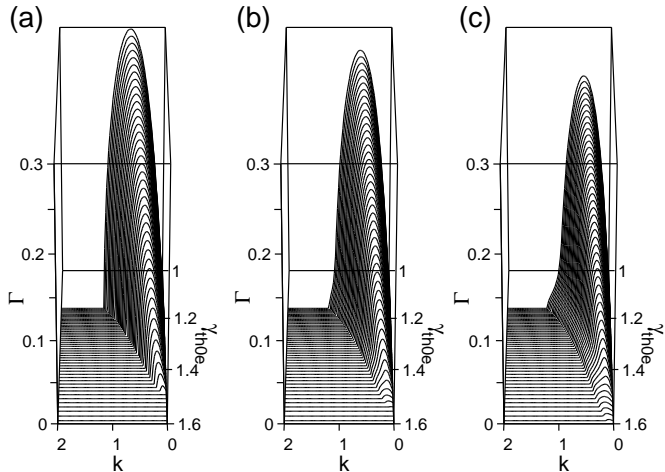,width=9cm}
\caption{Growth rate of the Weibel instability for all wavenumbers as function of the electron thermal Lorentz factor for shells having different velocity ($p_{x0}=2$, $V=0.75$), same electron density ($\eta=1$), same temperature, for different mixtures of plasma components: (a) $\alpha_{1}=1$, (b) $\alpha_{1}=0.75$, (c) $\alpha_{1}=0.5$.
\label{Fig. 3}}
\end{figure}

\begin{figure}
\epsfig{file=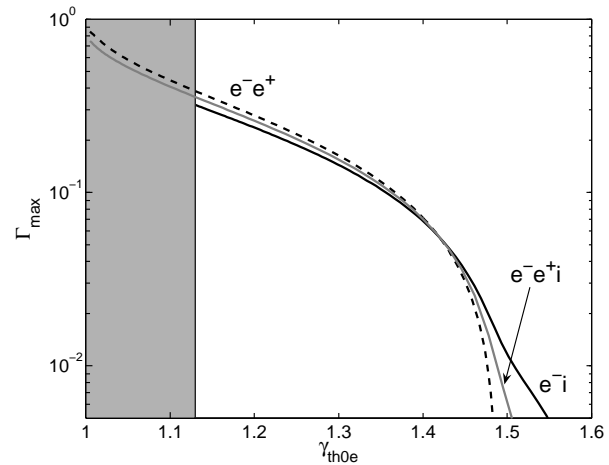,width=8cm}
\caption{Dependence of the maximum growth rate of the Weibel instability on the electron thermal Lorentz factor for shells having different velocity ($p_{x0}=2$, $V=0.75$), same electron density ($\eta=1$), same temperature, for different mixtures of plasma components: $\alpha_{1}=1$ (dashed line), $\alpha_{1}=0.75$ (solid gray line), $\alpha_{1}=0.5$ (solid black line). The light gray area indicates the forbidden area to guarantee the momentum conservation for $\alpha_{1}=0.5$.
\label{Fig. 4}}
\end{figure}

As for the analysis of shells having same temperature but different velocity and electron density, we consider the shell-1, moving with $p_{x0}=2$ ($\Rightarrow\gamma_{int}\approx4$), to be faster than the shell-2 (i.e. $V<1$). Among the scenarios we can examine by varying $V$ and $\eta$, let us study a case with shells having the same electron density, such that $\eta=1$. Since $V<1$, space charge effects are automatically taken into account because of (\ref{eq:19}). As the previous situation without space charge effects, a temperature exists above which the Weibel instability in the electron-positron plasma cannot develop anymore, whereas when ions are included, they lead to a small but non-negligible growth rate. This is shown in Fig. \ref{Fig. 3}, where the growth rate for all wavenumbers is plotted as a function of the electron thermal Lorentz factor for $V=0.75$ and $\eta=1$ kept fixed. The corresponding area of unstable points for this set of parameters is the gray filled area in Fig. \ref{Fig. 5}. In order to guarantee momentum conservation, we have chosen three values of $\alpha_1=1,0.75,0.5$, considering that for $\alpha=0.5$ the smallest  electron thermal Lorentz factor must be $\gamma_{th0e}=1.13$. The role played by baryons is illustrated more clearly in Fig. \ref{Fig. 4}, where the maximum growth rate of the Weibel instability is plotted against the electron thermal Lorentz factor (the light gray area is the forbidden area to guarantee momentum conservation for $\alpha=0.5$). By comparing Fig. \ref{Fig. 1} and Fig. \ref{Fig. 4} for the scenarios with $\alpha_1=1$ and $\alpha_1=0.5$, even though the inclusion of space charge effects leads to a less robust instability (i.e., as the temperature increases, the instability shutdowns earlier) by increasing the threshold for its occurrence, the baryon loading is always crucial to the occurrence of the instability for large temperatures. 

\section{Conclusions}

In summary, we have analyzed the possible role of the Weibel instability in magnetic field generation in GRB internal shocks. Starting from relativistic kinetic theory, we have studied the dynamic of two counter-streaming electron-positron-ion unmagnetized plasma shells with zero net charge. We have examined the Weibel instability in a hot plasma with arbitrary mixtures of plasma components and we have taken into account the space charge effects, so the mode is not purely electromagnetic and transverse ({\boldmath$k$}${\cdot}${\boldmath$E$}$\not=0$). Our discussion has been focused on the growth rate $\Gamma$ of the Weibel instability, since the maximum value of the magnetic field, which corresponds to the saturation level $b_{sat}$, is related to $\Gamma$ according to $b_{sat}\propto\Gamma^2$. Our calculations are performed in the CM reference frame, which means that the total linear momentum of the two shell system in this frame is zero.

Thanks to an analysis of the threshold for the occurrence of the instability for shells at the same temperature, it is possible to locate unstable scenarios by varying shell velocities and mixture component densities, taking momentum conservation into account. For shells with same velocity, temperature and electron density (i.e. no space charge effects), we have found that, even if the thermal spread slows down the Weibel instability and eventually shutdowns the instability, the presence of ions guarantees the occurrence of the instability for larger temperatures than the shutdown temperature for mixtures of leptons only. In the case of shells having different velocity, different electron density and same temperature, the occurrence of the Weibel instability for large temperatures and consequently the magnetic field generation, are still guaranteed by the baryon loading, even when the space charge effects are included. Baryons lead again to a small but non-negligible growth rate over the shutdown temperature for which the electron-positron plasma cannot allow anymore the occurrence of the Weibel instability. Preliminary PIC simulations qualitatively confirm our theoretical results concerning the growth rate and the threshold condition, and will be presented elsewhere.

\section*{ACKNOWLEDGMENTS}

Some of the authors (MF and LOS) acknowledge useful discussions with Prof. M. V. Medvedev. This work was partially supported by FCT (Portugal) through grants No. PDCT/FP/FAT/50190/2003 and No. POCI/FIS/55905/2004. The work of MF is supported by the FCT grant No. SFRH/BD/17678/2004.

\label{lastpage}

\end{document}